\documentclass[final]{aipproc}
\layoutstyle{6x9}%
\usepackage{amsfonts}
\usepackage{amsmath}
\usepackage[applemac]{inputenc}

\begin{document}

\title{On the balance equations for a dilute binary mixture in special relativity}


\author{Valdemar Moratto}{
  address={Departamento de F\'isica, Universidad Aut\'onoma Metropolitana-Iztapalapa, M\'exico D.F., M\'exico.}
}

\author{A. L. Garc\'ia-Perciante}{
  address={Departamento de Matem\'aticas Aplicadas y Sistemas, Universidad Aut\'onoma Metropolitana-Cuajimalpa, M\'exico D.F., M\'exico.}
}

\author{L. S. Garc\'ia-Col\'in}{
  address={Departamento de F\'isica, Universidad Aut\'onoma Metropolitana-Iztapalapa, M\'exico D.F., M\'exico.}
  ,altaddress={El Colegio Nacional, M\'exico D.F., M\'exico.} 
}

\begin{abstract}
 In this work we study the properties of a relativistic mixture of two non-reacting species in thermal local equilibrium. We use the full Boltzmann equation (BE) to find the general balance equations. Following conventional ideas in kinetic theory, we use the concept of chaotic velocity. This is a novel approach to the problem. The resulting equations will be the starting point of the calculation exhibiting the correct thermodynamic forces and the corresponding fluxes; these results will be published elsewhere.
\end{abstract}

\maketitle

\section{Introduction}

The purpose of this work is to obtain the balance equations for a system composed of two inert, non degenerate, dilute species. The calculation will be
carried out within the framework of special relativity. Its importance today relies on the fact that many applications can be found in a variety of problems,
from astrophysical systems to relativistic heavy ion collisions. Contrary to most of the approaches which are given in the literature for relativistic gases we here introduce, as in classical kinetic theory, the concept of chaotic velocity. This will be reflected on the fact that the resulting fluxes will be analogue to those obtained more than one hundred years ago by Maxwell and Clausius \cite{Maxwell,Clausius,Brush}, for whom the concept of chaotic fluxes was key in describing dissipative effects.
\\

In particular, we here address the first stage of this program namely, the derivation of the balance equations for the state variables. In our case these are
chosen to be, the number densities of both species, the baricentric velocity of the mixture and the internal energy density. Further the fluxes will be calculated in an arbitrary frame as the transformed dissipative fluxes which are identified in the local co-moving frame. This choice will reflect the importance of the Lorentz transformations in identifying the concept of chaotic velocity.
\\

\section{TRANSPORT EQUATIONS FOR THE NON-RELATIVISTIC BINARY
MIXTURE}

Kinetic theory constitutes a microscopic formalism from which the macroscopic properties of a system can be obtained based on the evolution of a distribution function for the molecules in the gas. For a binary mixture one considers
two separate distribution functions, one for each species, which satisfy two coupled Boltzmann equations, namely

\begin{eqnarray}
\frac{\partial f_{a}}{\partial t}+\textbf{v}_{a}\cdot\frac{\partial f_{a}}{\partial\textbf{r}}=J(f_{a}f_{a})+J(f_{a}f_{b})\label{Boltzmann Equation}
\end{eqnarray}
where the collision term is given by

\begin{eqnarray}J(f_{i}f_{j})=\int\cdots\int\left[f(\textbf{v}_{i}') f(\textbf{v}_{j}')-f(\textbf{v}_{i})f(\textbf{v}_{j})\right] \times\sigma\left(\textbf{v}_{i}\textbf{v}_{j}\rightarrow\textbf{v}_{i}' \textbf{v}_{j}'\right)
g_{ij}{\rm d}\textbf{v}_{j}{\rm d}\textbf{v}_{i}' {\rm d}\textbf{v}_{j}',
\end{eqnarray}
for each species $i,j=a,b$. Notice that the coupling of both integrodifferential
equations is given by the second term on the right hand side of Eq. (\ref{Boltzmann Equation}) namely,
the cross-collision term. The primes denote the values of $\textbf{v}_i$ after the binary collision
takes place, $\sigma\left(\textbf{v}_{i}\textbf{v}_{j}\rightarrow\textbf{v}_{i}' \textbf{v}_{j}'\right)
{\rm d}\textbf{v}_{i}' {\rm d}\textbf{v}_{j}'$ is the cross section, namely, the number
of molecules per unit time of species $i$ colliding with a molecule of species $j$
such that after the collision the molecules have velocities $\textbf{v}'_i$ in the range ${\rm d}\textbf{v}'_i$
and $\textbf{v}'_j$ in the range ${\rm d}\textbf{v}'_i$; $g_{ij}\equiv|\textbf{v}_i-\textbf{v}_j| =|\textbf{v}'_i-\textbf{v}'_j|$. We also recall the reader that the cross section $\sigma$ satisfies the principle of
microscopic reversibility, namely, it is invariant upon spatial and temporal reflections, so that,
\begin{eqnarray}
\sigma\left(\textbf{v}_{i}\textbf{v}_{j}\rightarrow\textbf{v}_{i}' \textbf{v}_{j}'\right)=\sigma\left(\textbf{v}_{i}'\textbf{v}_{j}'\rightarrow\textbf{v}_{i} \textbf{v}_{j}\right)\hspace{.5cm}{\rm{for}}\hspace{.5cm}i,j=a,b
\end{eqnarray}
thus guaranteeing the existence of inverse collisions.
\\

Since the treatment in both kinetic equations is symmetric we will only refer to species $a$. There is an analogous procedure for species $b$. In order to obtain the balance equations, the solution of the Boltzmann equation is not required since it already contains the information about the conservation of macroscopic quantities through the invariance of mass, momentum and
energy in each individual collision. As usual, we define the local particle density as,
\begin{eqnarray}
n_{a}=\int f_{a}{\rm{d}}\textbf{v}_{a},
\end{eqnarray}
\\
To obtain the conservation equations, one multiplies Eq. (\ref{Boltzmann Equation}) by the collision invariants and integrates over velocity space. For the mass conservation, we multiply by 1 and obtain
\begin{eqnarray}
\frac{\partial \rho_a}{\partial t} +\nabla\cdot\left(\rho_a \textbf u\right)=-\nabla\cdot \textbf{J}_a
\end{eqnarray}
where $\rho_a=m_an_a$ such that $\rho=\rho_a+\rho_b$ is the total density. The baricentric velocity $\textbf u$ is defined as
\begin{eqnarray}
n\textbf u=\sum^{b}_{i=a} n_i \textbf u_i=\sum^{b}_{i=a}\int\textbf v_i f_i {\rm d} \textbf{v}_i,\label{velocidad baricentrica clasica}
\end{eqnarray}
and the mass flux,
\begin{eqnarray}
\textbf{J}_a=m_{a}\int\textbf{k}_{a}f_{a}{\rm d}\textbf{k}_{a},\label{flujo de masa}
\end{eqnarray}
where the chaotic velocity for species $a$ is given by
\begin{eqnarray}
\textbf{k}_a=\textbf{v}_a-\textbf{u},\label{velocidad caotica}
\end{eqnarray}
and the flux satisfies
\begin{eqnarray}
\textbf{J}_a=-\textbf{J}_b.\label{difusion}
\end{eqnarray}
\\

For the momentum balance equation, we consider the collisional invariant $m_a \textbf {v}_a$. With the help of $\textbf{k}_a=\textbf{v}_a-\textbf{u}$, one obtains
\begin{eqnarray}
\frac{\partial}{\partial t}(\rho\textbf{u}) =-\nabla\cdot(\tau+\rho\textbf{u}\textbf{u}),
\end{eqnarray}
where
\begin{eqnarray}
\tau=\sum_{i=a}^{b} m_{i}\int\textbf{k}_{i}\textbf{k}_{i}f_{i}{\rm d} \textbf{k}_{i}\label{tensor de esfuerzos}
\end{eqnarray}
is the stress tensor.
\\

Finally for the energy balance equation we use the invariance of $\frac{1}{2}m_a |\textbf{v}_a|^2$, and in a similar fashion we obtain,
\begin{eqnarray}
\frac{\partial}{\partial t}(\rho e) =-\nabla\cdot\left(\rho e\textbf{u} +\textbf{J}_{q}\right) -\tau:\nabla\textbf{u},
\end{eqnarray}
where
\begin{eqnarray}
\textbf {J}_q=\sum^{b}_{i=a} \textbf J_{q_i} =\sum^{b}_{i=a} \frac{m_i}{n_i}\int\textbf k_i k_i f_i {\rm d} \textbf{k}_i,\label{flujo de calor}
\end{eqnarray}
is the heat flux, and the local internal energy is given by
\begin{eqnarray}
\rho e=\sum^{b}_{i=a} \frac{1}{2} m_i \int k_i^2 f_i{\rm d} \textbf{k}_i,\label{energia interna}
\end{eqnarray}
where $k_i$ is the magnitude of $\textbf{k}_i$.
\\

It is important to emphasize the fact that the averages in Eqs. (\ref{flujo de masa}), (\ref{tensor de esfuerzos}) and (\ref{flujo de calor}) are calculated with the chaotic part of the molecular velocity $\textbf{k}_a$. Isolating the chaotic part of $\textbf{v}_a$ leads to the identification of the dissipative fluxes. In fact, his is the physical meaning of heat firstly introduced by Maxwell \cite{Maxwell} \cite{Brush}.

\section{Relativistic Kinetic Theory in Special Relativity}

The system we consider, and that will refer to as relativistic binary mixture,
is a gas constituted of two non-degenerate species that do not react but only
interact through collisions. The gas is diluted but the molecular velocity of both
species is high enough for relativistic effects to be relevant. This is reflected in
the fact that the relativistic parameters $z_i = kT/m_ic^2$ are grater than one
namely, the thermal energy is larger than the rest energy of the molecules. Here
$T$ is the temperature of the gas, $k$ the Boltzmann constant, $c$ the speed of light
and $m_i$ the rest mass of each species. Additionally, we consider the system
in the absence of external forces.
Before addressing the kinetic theory of the mixture we review some basic
aspects of the relativistic kinetic theory. The relativistic Boltzmann equation
reads \cite{Israel,deGroot vanLeeuwen vanWeert,Kremer},
\begin{eqnarray}
v^{\alpha}f_{,\alpha}=J\left(ff\right)\label{Boltzmann equation RE}
\end{eqnarray}
where $f$ is the distribution function as before. Proofs of the invariance of this quantity are available in Refs. \cite{deGroot vanLeeuwen vanWeert} and \cite{Kremer}. Here $v^\alpha =\gamma_\omega (\overrightarrow\omega, c)$ is the molecular four-velocity
with $\gamma_\omega=\left(1-(\omega/c)^2\right)^{-1/2}$ being the Lorentz factor. Thus, the left hand side
of Eq. (\ref{Boltzmann equation RE}) is clearly an invariant. The collision term can also be written in an
invariant fashion as follows
\begin{eqnarray}
J(ff)
=\int\int\left(f'f_{1}'-ff_{1}\right)
\mathcal{F}\sigma \left(\Omega\right) {\rm d} \Omega  {\rm d} v^*_1,\label{collisional term}
\end{eqnarray}
where $ {\rm d} v^*_1=\frac{{\rm d}^3v_1}{v^4_1}$ and $\mathcal{F}$ are Lorentz invariants. The latter is known as the invariant flux and is given by \cite{Kremer},
\begin{eqnarray}
\mathcal{F}
=\frac{1}{c^2}v^4v^4_1=\frac{1}{c}\sqrt{\left(v^\nu v_{1\nu}\right)^2-c^2}
=\frac{1}{c}\sqrt{\left(\gamma_\omega \gamma_{\omega 1}\left(\overrightarrow\omega\cdot\overrightarrow\omega_1-c^2\right)\right)^2-c^4}
\end{eqnarray}
which reduces to the relative velocity in the non-relativistic limit.
\\

It is well known that without a solution of Boltzmann equation we can obtain two results, the H theorem and the balance equations. Since in this work we focus on the balance equations we will not discuss any
further the properties of the Boltzmann equation or its methods of solution.
\\

As in the non-relativistic case, the Boltzmann equation is multiplied by each of
the collision invariants and integrated to yield the transport equations. However, the dissipative fluxes cannot be clearly identified if the chaotic velocity is not introduced explicitly. As extensively discussed in Ref. \cite{Microscopic nature of dissipative effects ANA-ALFREDO}, Lorentz
transformations can be used in order to write the molecular velocity measured by an arbitrary observer in terms of the chaotic velocity which is the one measured in a local frame that moves with the fluid element namely, the co-moving frame. This is expressed as
\begin{eqnarray}
v^\mu=\mathcal{L}^\mu_\nu K^\nu\label{Lorentz Transformation}
\end{eqnarray}
where $K^\nu=\gamma_k\left(\overrightarrow k,c\right)$ is the chaotic velocity. The establishment of the particle
four-flux and energy-momentum tensor in this context for the one-component
system is discussed in Ref. \cite{Microscopic nature of dissipative effects ANA-ALFREDO}. In the next section we will follow these ideas
in order to obtain the balance equations for the relativistic mixture.

\section{balance equations for the relativistic mixture}

The Boltzmann equations for the special relativistic binary mixture are
\begin{eqnarray}
v_{1}^{\alpha}f_{1,\alpha}=J\left(f_{1}f_{1}\right) +J\left(f_{1}f_{2}\right)\label{Boltzmann equation Species 1}
\end{eqnarray}
\begin{eqnarray}
v_{2}^{\alpha}f_{2,\alpha}=J\left(f_{2}f_{2}\right) +J\left(f_{2}f_{1}\right)\label{Boltzmann equation Species 2}
\end{eqnarray}
where $J(f_if_j)$ is defined in the same way as in Eq (\ref{collisional term}) and the indices 1 and 2 indicate species.
\\

As before, in order to obtain the balance equations one multiplies Eqs. (\ref{Boltzmann equation Species 1}) and (\ref{Boltzmann equation Species 2}) by the corresponding collision invariants, in this case $m_i$ and $v^\mu_i$, $i=1,2$ and integrates over velocity space. By multiplying Eq. (\ref{Boltzmann equation Species 1}) by $m_1$ and integrating over the Lorentz invariant element ${\rm d} v^*_1$ one finds,
\begin{eqnarray}
N_{1;\mu}^{\mu}=0\label{particle conservation}
\end{eqnarray}
where
\begin{eqnarray}
N^\mu=N_1^\mu+N_2^\mu =\int v_{1}^{\mu}f_{1}{\rm d} v^*_1+\int v_{2}^{\mu} f_{2} {\rm d} v^*_2,\label{tetraflux}
\end{eqnarray}
which corresponds to the four-flux of particles in an arbitrary frame. Equation (\ref{tetraflux}) leads to the definition of a baricentric velocity given by $N^\mu=nU^\mu$, where $n=n_1+n_2$. Notice the similarity with the definition in the non relativistic case, Eq. (\ref{velocidad baricentrica clasica}). By using the transformation in Eq. (\ref{Lorentz Transformation}) one can write
\begin{eqnarray}
N_1^\mu=\mathcal{L}^\mu_\nu J_1^\nu,
\end{eqnarray}
where $J_1^\nu$ is the dissipative particle four-flux in the co-moving frame and it satisfies $J_1^\nu+J_2^\nu=(\vec{0},n)$.
Introducing this relation in Eq. (\ref{particle conservation}) one finds after some laborious algebra, that
\begin{eqnarray}
\left(\mathcal{L}_{a}^{b}J_{1}^{a}\right)_{,b}
+\frac{1}{c^{2}}\left[\gamma_{u}\frac{\partial} {\partial t}U_l-U_l\frac{u}{c^2} \gamma_{u}^{3}\frac{\partial u}{\partial t}\right]J_{1}^{l}
+\frac{1}{c^{2}} \gamma_{u}U_l\frac{\partial}{\partial t}J_{1}^{l}+n_{1}\theta+U^{\mu}n_{1,\mu}=0,\label{PARTICLE CONS COMPLETE}
\end{eqnarray}
where $U^\nu=\gamma_u\left(\vec{u},c\right)$, $\gamma_u=\left(1-(u/c)^2\right)^{-1/2}$ and $\theta=U^\alpha_{,\alpha}$. Equation (\ref{PARTICLE CONS COMPLETE}),
which corresponds to the mass conservation equation for species 1; latin indices correspond to the spatial part of any tensor and run form 1 to 3.
Notice that in Euler's regime only the last two terms do not vanish.
\\

For the energy momentum conservation we multiply Eq. (\ref{Boltzmann equation Species 1}) by $v_1^\mu$, integrate on velocity space and obtain,
\begin{eqnarray}
T_{1;\nu}^{\mu\nu}=0\label{energy-momentum conservation}
\end{eqnarray}
where
\begin{eqnarray}
T^{\mu\nu}=T_{1}^{\mu\nu}+T_{2}^{\mu\nu} =m_{1}\int v_{1}^{\mu}v_{1}^{\nu}f_{1}{\rm d} v^*_1
+m_{2}\int v_{2}^{\mu}v_{2}^{\nu}f_{2}{\rm d} v^*_2,\label{Tmunu}
\end{eqnarray}
is the energy-momentum tensor.
\\

In order to explicitly calculate Eq. (\ref{Tmunu}) we start by noticing that the energy-momentum tensor, in the co-moving frame can be written as,
\begin{eqnarray}
\widetilde{T}^{\mu\nu}\ddot{=}\left(\begin{array}{cccc}
p & 0 & 0 & 0\\
0 & p & 0 & 0\\
0 & 0 & p & 0\\
0 & 0 & 0 & \epsilon\end{array}\right)
+\left(\begin{array}{cccc}
\pi_{11} & \pi_{12} & \pi_{13} & q_{1}\\
\pi_{12} & \pi_{22} & \pi_{23} & q_{2}\\
\pi_{13} & \pi_{23} & \pi_{33} & q_{3}\\
q_{1} & q_{2} & q_{3} & \pi_{44}\end{array}\right),\label{energy momentum tensor COMOVIL}
\end{eqnarray}
where first term corresponds to the equilibrium situation \cite{Weinberg}. Here the hydrostatic pressure $p$ and internal energy density $\epsilon$ are given by
\begin{eqnarray}
p=p_{1}+p_{2}=m_{1}\int K_{1}^{a}K_{1}^{a}f_{1}^{(0)} {\rm d}^{3}K_{1}^{*}+m_{2}\int K_{2}^{a}K_{2}^{a}f_{2}^{(0)} {\rm d}^{3}K_{2}^{*}
\end{eqnarray}
\begin{eqnarray}
\epsilon=n_1e_{1}+n_2e_{2}=m_{1}\int K_{1}^{4}K_{1}^{4} f_{1}^{(0)}{\rm d}^{3}K_{1}^{*}+m_{2}\int K_{2}^{4}K_{2}^{4} f_{2}^{(0)}{\rm d}^{3}K_{2}^{*},
\end{eqnarray}
where $f_1^{(0)}$ denotes the local equilibrium solution to Eq. (\ref{Boltzmann equation Species 1}). This function is the well known J\"uttner distribution function \cite{Juttner},\cite{deGroot vanLeeuwen vanWeert}\cite{Kremer}.
\\

The second term in Eq. (\ref{energy momentum tensor COMOVIL}) corresponds to the non equilibrium situation, where we identify,
\begin{eqnarray}
q^{a}=q^{a}_1+q^{a}_2 =c^{2}\int\gamma_{k_{1}}K_{1}^{a}f_{1} {\rm d}K_{2}^{*}+c^{2} \int\gamma_{k_{2}}K_{2}^{a} f_{2}{\rm d}K_{2}^{*}\label{heat flux RE}
\end{eqnarray}
and
\begin{eqnarray}
\pi^{ab}=\pi^{ab}_{1}+\pi^{ab}_{2}=m_{1}\int K_{1}^{a}K_{1}^{b}f_{1} {\rm d}K_{1}^{*}+m_{2}\int K_{1}^{a}K_{2}^{b}f_{2} {\rm d}K_{2}^{*}.\label{viscous tensor RE}
\end{eqnarray}
In Ref. \cite{Microscopic nature of dissipative effects ANA-ALFREDO} it is shown that
\begin{eqnarray}
\pi_{44}=0
\end{eqnarray}
by introducing a Chapman and Enskog expansion. The quantities in Eq. (\ref{heat flux RE}) and (\ref{viscous tensor RE}) are identified as the heat flux and viscous tensor since they correspond to the average of the chaotic energy and momentum fluxes respectively. This is based on the physical interpretation of dissipative fluxes as is mentioned in Refs. \cite{Maxwell}\cite{Clausius}\cite{Brush}.

We then proceed in the same fashion as in the particle conservation, by writing
\begin{eqnarray}
T^{\mu\nu}=\mathcal{L}_\alpha^\mu\mathcal{L}_\beta^\mu\tilde{T}^{\alpha\beta}
\end{eqnarray}
which yields
\begin{eqnarray}
T^{\alpha\beta}= pg^{\alpha\beta}+\frac{1}{c^{2}} \left(p+\epsilon\right)U^{\alpha}U^{\beta} +\frac{1}{c^{2}} \left(U^{\alpha}\mathcal{L}_{a}^{\beta}q^{a}+U^{\beta} \mathcal{L}_{a}^{\alpha}q^{a}\right) +\mathcal{L}_{a}^{\beta}\mathcal{L}_{b}^{\alpha} \Pi^{ab},\label{energy momentum tensor FINAL}
\end{eqnarray}
the first two terms correspond to the equilibrium case, while the second term represents the contribution of the heat flux and the third term includes the viscosities. Here
\begin{eqnarray}\nonumber
\Pi^{ab}{=}\left(\begin{array}{cccc}
\pi_{11} & \pi_{12} & \pi_{13} & 0\\
\pi_{12} & \pi_{22} & \pi_{23} & 0\\
\pi_{13} & \pi_{23} & \pi_{33} & 0\\
0 & 0 & 0 & 0\end{array}\right).
\end{eqnarray}
\\

The last step is to work out the derivative in Eq. (\ref{energy-momentum conservation}) with the help of Eq. (\ref{energy momentum tensor FINAL}). This leads to
\begin{eqnarray}
\frac{1}{c^{2}}\left[U^{\beta}e U^\mu n_{,\mu}+nU^{\beta}U^\mu e_{,\mu}+neU^{\beta}U^{\nu}_{,\nu}+neU^\mu{U}_{,\mu}^{\beta}\right]
\label{MOMENT CONS COMPLETE}\nonumber\\ +p_{,\alpha}h^{\beta\alpha}+\frac{p}{c^{2}}\left(U^{\beta}U^{\nu}_{,\nu}+U^\mu{U}_{,\mu}^{\beta}\right)\\ +\frac{1}{c^{2}}\left[\left(U^{\alpha}\mathcal{L}_{a}^{\beta}\right)_{,\alpha}q^{a}+\left(U^{\alpha}\mathcal{L}_{a}^{\beta}\right)q_{,\alpha}^{a} +\left(U^{\beta}\mathcal{L}_{a}^{\alpha}\right)_{,\alpha}q^{a}+\left(U^{\beta}\mathcal{L}_{a}^{\alpha}\right)q_{,\alpha}^{a}\right]\nonumber\\
+\left(\mathcal{L}_{a}^{\beta}\mathcal{L}_{b}^{\alpha}\right)_{,\alpha}{\Pi}^{ab} +\left(\mathcal{L}_{a}^{\beta} \mathcal{L}_{b}^{\alpha}\right){\Pi}_{,\alpha}^{ab}
=0,\nonumber
\end{eqnarray}
which is the energy momentum balance equation for the mixture. In Euler's regime it reduces to,
\begin{eqnarray}
\frac{1}{c^{2}}U^{\beta}\left[n\dot{e}+p\theta\right]
+\frac{1}{c^{2}}\left(ne+p\right)\dot{U}^{\beta} +p_{,\alpha}h^{\beta\alpha}=0,\label{T coma}
\end{eqnarray}
where the notation $\dot{(\hspace{.2cm})}=U^\mu(\hspace{.2cm})_{,\mu}$ and $\theta=U^{\mu}_{,\mu}$ has been introduced.
To isolate the energy balance, we calculate $U_{\mu}T_{;\nu}^{\mu\nu}$ which, again in Euler's regime, reads
\begin{eqnarray}
-\left(n\dot{e}+p\theta\right)=0.
\end{eqnarray}
Introducing the previous equation in Eq. (\ref{T coma}) we obtain
\begin{eqnarray}
\frac{1}{c^{2}}\left(ne+p\right)\dot{U}^{\beta}+h^{\beta\nu}p_{,\nu}=0.
\end{eqnarray}
It is important stress that the form of Eq. (\ref{energy momentum tensor FINAL}) has been obtained form microscopic grounds.

\section{Conclusions}

We have obtained the balance equations for a relativistic, diluted, non degenerate, mixture based solely on microscopic grounds. We explicitly used the concept of chaotic velocity to identify the dissipative fluxes. The results are shown in Eqs. (\ref{PARTICLE CONS COMPLETE}) and (\ref{MOMENT CONS COMPLETE}) from which the Euler equations are recovered in the equilibrium case. The dissipative terms in these equations differ from the ones obtained following the standard procedure (see appendix B of Ref. \cite{Microscopic nature of dissipative effects ANA-ALFREDO}).
\\

In particular, the Euler equations for the mixture are required in order to solve the linearized Boltzmann equation within the Chapman and Enskog expansion. This constitutes work in progress and will be publish elsewhere.
\\




\bibliographystyle{aipproc}   


\end{document}